\documentclass[11pt]{article}
\usepackage{amssymb,amsmath,amsfonts}
\usepackage{graphicx}
\usepackage{graphics}
\usepackage{eepic,epsfig}

\begin{document}

\title{Induced vacuum currents in anti-de Sitter space \\
with toral dimensions}
\author{E. R. Bezerra de Mello$^{1}$\thanks{
E-mail: emello@fisica.ufpb.br}, A. A. Saharian$^{2,1}$\thanks{%
E-mail: saharian@ysu.am}, V. Vardanyan$^{2,3}$ \vspace{0.3cm} \\
\textit{$^{1}$Departamento de F\'{\i}sica, Universidade Federal da Para\'{\i}%
ba}\\
\textit{58.059-970, Caixa Postal 5.008, Jo\~{a}o Pessoa, PB, Brazil}\vspace{%
0.3cm}\\
\textit{$^2$Department of Physics, Yerevan State University,}\\
\textit{1 Alex Manoogian Street, 0025 Yerevan, Armenia} \vspace{0.3cm}\\
\textit{$^3$Institut f\"{u}r Theoretische Physik, Ruprecht-Karls-Universit%
\"{a}t Heidelberg,}\\
\textit{Philosophenweg 16, 69120 Heidelberg, Germany}}
\maketitle

\begin{abstract}
We investigate the Hadamard function and the vacuum expectation
value of the current density for a charged massive scalar field on a
slice of anti-de Sitter (AdS) space described in Poincar\'{e}
coordinates with toroidally compact dimensions. Along compact
dimensions periodicity conditions are imposed on the field with
general phases. Moreover, the presence of a constant gauge field is
assumed. The latter gives rise to Aharonov-Bohm-like effects on the
vacuum currents. The current density along compact dimensions is a
periodic function of the gauge field flux with the period equal to
the flux quantum. It vanishes on the AdS boundary and, near the
horizon, to the leading order, it is conformally related to the
corresponding quantity in Minkowski bulk for a massless field. For
large values of the length of the compact dimension compared with
the AdS curvature radius, the vacuum current decays as power-law for
both massless and massive fields. This behavior is essentially
different from from the corresponding one in Minkowski background,
where the currents for a massive field are suppressed exponentially.
\end{abstract}

\bigskip


\bigskip

\section{Introduction}

\label{sec:introd}

In quantum field theory on curved backgrounds, the anti-de Sitter (AdS)
spacetime is interesting from several points of view. The corresponding
metric tensor is a maximally symmetric solution of the Einstein equations in
the presence of a negative cosmological constant and, because of the high
degree of symmetry, numerous physical problems are exactly solvable on its
background. These investigations may help to gain deeper understanding of
the influence of gravity on quantum matter. The early interest to the
dynamics of quantum fields in AdS spacetime was motivated by principal
questions of the quantization on curved backgrounds. The presence of both
regular and irregular modes and the possibility of getting an interesting
causal structure, lead to a number of interesting features that have no
analogs in Minkowski spacetime field theory. Being a constant negative
curvature manifold, AdS space provides a convenient infrared regulator in
interacting quantum field theories \cite{Call90}, cutting infrared
divergences without reducing the symmetries. The importance of AdS spacetime
as a gravitational background increased after the discovery that it
generically arises as a ground state in extended supergravity and in string
theories, and also as the near horizon geometry of the extremal black holes
and branes.

More recent interest in this subject was generated in connection with two
types of models where AdS geometry plays a crucial role. The first one,
called AdS/CFT correspondence \cite{Mald98} (for a review see \cite{Ahar00}%
), represents a realization of the holographic principle and provides a
mapping between string theories or supergravity in the AdS bulk and a
conformal field theory living on its boundary. In this mapping the solutions
in AdS space play the role of classical sources for the correlators in the
boundary field theory. The AdS/CFT correspondence has many interesting
consequences and provides an opportunity to investigate non-perturbative
field-theoretical effects at strong couplings. Among the recent developments
of such a holography is the application to strong-coupling problems in
condensed matter physics (holographic superconductors, quantum phase
transitions, topological insulators). The second type of model with AdS
spacetime as a background geometry is a realization of a braneworld scenario
with large extra dimensions, and provides a solution to the hierarchy
problem between the gravitational and electroweak mass scales (for reviews
see \cite{Ruba01}). Braneworlds naturally appear in string/M-theory context
and give a novel setting for discussing phenomenological and cosmological
issues related to extra dimensions, including the problem of the
cosmological constant.

The presence of extra compact dimensions is an inherent feature of all these
models. In quantum field theory, the boundary conditions imposed on the
field operator along compactified dimensions lead to a number of interesting
physical effects that include topological mass generation, instabilities in
interacting field theories and symmetry breaking. These boundary conditions
modify the spectrum of the zero-point fluctuations, as a result the vacuum
energy density and the stresses are changed. This is the well-known
topological Casimir effect. This effect has been investigated for large
number of geometries and has important implications on all scales, from
mesoscopic physics to cosmology (for reviews see \cite{Most97}). The vacuum
energy depends on the size of extra dimensions and this provides a
stabilization mechanism for moduli fields in Kaluza-Klein-type models and in
braneworld scenario. In particular, motivated by the problem of radion
stabilization in Randall-Sundrum-type braneworlds, the investigations of the
Casimir energy on AdS bulk have attracted a great deal of attention\footnote{%
See references in \cite{Eliz13}.}. The Casimir effect in AdS spacetime with
compact internal spaces has been considered in \cite{Flac03}. The vacuum
energy generated by the compactification of extra dimensions can also serve
as a model of dark energy needed for the explanation of the present
accelerated expansion of the universe.

An important characteristic associated with charged fields is the vacuum
expectation value (VEV) of the current density. Although the corresponding
operator is local, because of the global nature of the quantum vacuum, this
expectation value carries information about both the geometry and topology
of the background space. Moreover, this VEV acts as the source in the
semiclassical Maxwell equations and therefore plays an important role in
modeling a self-consistent dynamics involving the electromagnetic field. The
VEV of the current density for a fermionic field in flat spaces with toral
dimensions has been investigated in \cite{Bell10}. Applications were given
to the electrons of a graphene sheet rolled into cylindrical and toroidal
shapes and described in terms of an effective Dirac theory in a
two-dimensional space. Combined effects of the compactification and
boundaries are discussed in \cite{Bell13}. The finite temperature effects on
the current densities for scalar and fermionic fields in topologically
nontrivial spaces have been studied in \cite{Beze13b}. The VEV of the
current density for charged scalar and Dirac spinor fields in de Sitter
spacetime with toroidally compact spatial dimensions are considered in \cite%
{Bell13b}. The effects of nontrivial topology induced by the
compactification of a cosmic string along its axis have been discussed in
\cite{Beze13}.

In the present paper we investigate the VEV of the current density for a
charged scalar field in a slice of AdS spacetime covered by Poincar\'{e}
coordinates assuming that a part of spatial coordinates are compactified to
a torus. In addition to the background gravitational field, we also assume
the presence of a constant gauge field interacting with the field. Though
the corresponding field strength vanishes, the nontrivial spatial topology
gives rise Aharonov-Bohm-like effect on the current density along compact
dimensions. This current is a source of magnetic fields in the
uncompactified subspace, in particular, on the branes in braneworld
scenario. The problem under consideration is also of separate interest as an
example with gravitational and topological polarizations of the vacuum for
charged fields, where one-loop calculations can be performed in closed form..

The outline of the paper is as follows. In the next section we describe the
geometry of the problem and evaluate the Hadamard function for a charged
massive scalar field obeying general quasiperiodicity conditions along
compact dimensions. By using this function, in section \ref{sec:Current},
the VEV of the current density is investigated. The behavior of this VEV in
various asymptotic regions of the parameters is discussed. The main results
of the paper are summarized in section \ref{sec:Conc}. The main steps for
the transformation of the Hadamard function to its final expression are
described in Appendix.

\section{Hadamard function}

\label{sec:Hadam}

As a background geometry we consider $(D+1)$-dimensional AdS spacetime. In
Poincar\'{e} coordinates the corresponding line element is expressed as
\begin{equation}
ds^{2}=g_{\mu \nu }dx^{\mu }dx^{\nu }=e^{-2y/a}\eta _{ik}dx^{i}dx^{k}-dy^{2},
\label{metric}
\end{equation}%
where $\eta _{ik}=\mathrm{diag}(1,-1,\ldots ,-1)$, $i,k=0,1,\ldots ,D-1$, is
the metric tensor for $D$-dimensional Minkowski spacetime and $-\infty
<y<+\infty $. The parameter $a$ is the AdS curvature radius and is related
with the cosmological constant. Note that the Poincar\'{e} coordinates cover
a part of the AdS manifold and there is a horizon corresponding to the
hypersurface $y=+\infty $. In what follows we assume that the coordinates $%
x^{l}$, with $l=p+1,\ldots ,D-1$, are compactified to circles with the
lengths $L_{l}$, so $0\leqslant x^{l}\leqslant L_{l}$. For the coordinates $%
x^{l}$, with $l=1,2,\ldots ,p$, one has $-\infty <x^{l}<+\infty $. Hence,
the subspace perpendicular to the $y$-axis has a topology $R^{p}\times
(S^{1})^{q}$, where $q+p=D-1$. The coordinates in uncompactified and
compactified subspaces will be denoted by $\mathbf{x}_{p}=(x^{1},\ldots
,x^{p})$ and $\mathbf{x}_{q}=(x^{p+1},\ldots ,x^{D-1})$, respectively.

\begin{figure}[tbph]
\begin{center}
\epsfig{figure=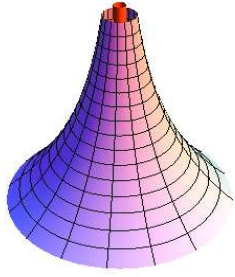,width=5cm,height=5cm}
\end{center}
\caption{The spatial section of the geometry under consideration for $D=2$
embedded into a 3-dimensional Euclidean space. }
\label{fig1}
\end{figure}

Introducing a new coordinate
\begin{equation}
z=ae^{y/a},\;0\leqslant z<\infty ,  \label{z}
\end{equation}%
the line element is presented in a conformally-flat form%
\begin{equation}
ds^{2}=(a/z)^{2}(\eta _{ik}dx^{i}dx^{k}-dz^{2}).  \label{metric2}
\end{equation}%
The hypersurfaces identified by $z=0$ and $z=\infty $ correspond to the AdS
boundary and horizon, respectively. In figure \ref{fig1} we have displayed
the spatial geometry corresponding to (\ref{metric}) for $D=2$ embedded into
the 3-dimensional Euclidean space. The compact dimension corresponds to the
circles. We have also displayed the flux of the gauge field strength which
threads the compact dimension (see below). Note that for a given $z$, the
proper length of the $l$th compact dimension is given by $L_{(p)l}=aL_{l}/z$
and it decreases with increasing $z$.

In this paper we are interested in the evaluation of the VEV of the current
density
\begin{equation}
j_{\mu }(x)=ie[\varphi ^{+}(x)D_{\mu }\varphi (x)-(D_{\mu }\varphi
^{+}(x))\varphi (x)],  \label{jmu}
\end{equation}%
associated with a massive charged scalar field, $\varphi (x)$, in the
presence of an external classical gauge field, $A_{\mu }$. In (\ref{jmu}), $%
D_{\mu }=\nabla _{\mu }+ieA_{\mu }$, with $\nabla _{\mu }$ being the
standard covariant derivative. The corresponding equation of motion is given
by
\begin{equation}
\left( g^{\mu \nu }D_{\mu }D_{\nu }+m^{2}+\xi R\right) \varphi (x)=0,
\label{fieldeq}
\end{equation}%
where $\xi $ is the curvature coupling parameter. The most important special
cases correspond to minimally and conformally coupled fields with the
parameters $\xi =0$ and $\xi =(D-1)/(4D)$, respectively. In the geometry
under consideration for the scalar curvature one has $R=-D(D+1)/a^{2}$. In
models with nontrivial topology one need also to specify the periodicity
conditions obeyed by the field operator along compact dimensions. Here we
consider quasiperiodicity conditions%
\begin{equation}
\varphi (t,\mathbf{x}_{p},\mathbf{x}_{q}+L_{l}\mathbf{e}_{l})=e^{i\alpha
_{l}}\varphi (t,\mathbf{x}_{p},\mathbf{x}_{q}),  \label{BC}
\end{equation}%
with constant phases $\alpha _{l}$ and with $\mathbf{e}_{l}$\ being the unit
vector along the $l$th compact dimension, $l=p+1,\ldots ,D-1$. The special
cases of untwisted and twisted fields correspond to $\alpha _{l}=0$ and $%
\alpha _{l}=\pi $, respectively.

In what follows we assume that the gauge field is constant, $A_{\mu }=%
\mathrm{const}$. Though the corresponding field strength vanishes, the
nontrivial topology of the background space gives rise to
Aharaonov-Bohm-like effects on the VEVs of physical observables. By a gauge
transformation, the problem with a constant gauge field may be reduced to
the problem in the absence of the gauge field with the shifted phases in the
periodicity conditions (\ref{BC}). Indeed, let us consider two sets of the
fields $(\varphi ,A_{\mu })$ and $(\varphi ^{\prime },A_{\mu }^{\prime })$
related by the gauge transformation $\varphi =\varphi ^{\prime }e^{-ie\chi }$%
, $A_{\mu }=A_{\mu }^{\prime }+\partial _{\mu }\chi $. Choosing $\chi
=A_{\mu }x^{\mu }$, we see that in the new gauge the vector potential
vanishes, whereas the field operator obeys the periodicity condition%
\begin{equation}
\varphi ^{\prime }(t,\mathbf{x}_{p},\mathbf{x}_{q}+L_{l}\mathbf{e}_{l})=e^{i%
\widetilde{\alpha }_{l}}\varphi ^{\prime }(t,\mathbf{x}_{p},\mathbf{x}_{q}),
\label{BC2}
\end{equation}%
with the new phases%
\begin{equation}
\widetilde{\alpha }_{l}=\alpha _{l}+eA_{l}L_{l}.  \label{alftilde}
\end{equation}%
In this expression, $A_{l}L_{l}=\oint dx^{l}A_{l}$ is the flux of the field
strength which threads the $l$th compact dimension. Note that in models with
compact extra dimensions the gauge field fluxes generate a potential for the
moduli fields providing a stabilization mechanism for them (for a review see
\cite{Doug07}). In the evaluation of the VEVs we can use both the sets $%
(\varphi ,A_{\mu })$ and $(\varphi ^{\prime },A_{\mu }^{\prime })$. In the
following, the evaluation procedure for the VEV of the current density is
based on the set $(\varphi ^{\prime },A_{\mu }^{\prime })$. The field obeys
the equation (\ref{fieldeq}) and the current density is given by (\ref{jmu})
with $D_{\mu }=\nabla _{\mu }$ in both the formulas. For simplicity of the
notations we shall omit the prime in the discussion below.

The VEV of the current density can be expressed in terms of the
Hadamard function
\begin{equation}
G^{(1)}(x,x^{\prime })=\langle 0|\varphi (x)\varphi ^{+}(x^{\prime
})+\varphi ^{+}(x^{\prime })\varphi (x)|0\rangle ,  \label{G1}
\end{equation}%
being $|0\rangle $ the vacuum state. Here we consider the vacuum
corresponding to the Poincar\'{e} coordinate system, the so called Poincar%
\'{e} vacuum. For the VEV of the current density one gets%
\begin{equation}
\langle 0|j_{\mu }(x)|0\rangle =\langle j_{\mu }\rangle =\frac{i}{2}%
e\lim_{x^{\prime }\rightarrow x}(\partial _{\mu }-\partial _{\mu }^{\prime
})G^{(1)}(x,x^{\prime }).  \label{jmu2}
\end{equation}%
Let us denote by $\{\varphi _{\sigma }^{(+)}(x),\varphi _{\sigma
}^{(-)}(x)\} $ the complete set of normalized positive- and negative-energy
solutions to the field equation obeying the periodicity condition (\ref{BC2}%
). The set of quantum numbers $\sigma $ that label the mode functions will
be specified below. Expanding the field operator in terms of the complete
set we obtain the following expression for the Hadamard function%
\begin{equation}
G^{(1)}(x,x^{\prime })=\sum_{\sigma }\sum_{s=\pm }\varphi _{\sigma
}^{(s)}(x)\varphi _{\sigma }^{(s)\ast }(x^{\prime }).  \label{G11}
\end{equation}

For the problem under consideration, we can choose the mode function in the
factorized form%
\begin{equation}
\varphi _{\sigma }^{(\pm )}(x)=e^{i\mathbf{k}_{p}\mathbf{x}_{p}+i\mathbf{k}%
_{q}\mathbf{x}_{q}\mp i\omega t}f(z),  \label{phipm}
\end{equation}%
where $\mathbf{k}_{p}=(k_{1},\ldots ,k_{p})$ and $\mathbf{k}%
_{q}=(k_{p+1},\ldots ,k_{D-1})$. For the components of the wave vector along
uncompactified dimensions one has $-\infty <k_{l}<+\infty $, $l=1,\ldots ,p$%
, and for the components along compact dimensions from the condition (\ref%
{BC2}) we find%
\begin{equation}
k_{l}=(2\pi n_{l}+\tilde{\alpha}_{l})/L_{l},\;l=p+1,\ldots ,D-1,  \label{kl}
\end{equation}%
with $n_{l}=0,\pm 1,\pm 2,\ldots $. Substituting (\ref{phipm}) into the
field equation we obtain the equation for the function $f(z)$. The solution
of the latter is presented in the form
\begin{equation}
f(z)=Cz^{D/2}Z_{\nu }(\lambda z),  \label{fy}
\end{equation}%
where $Z_{\nu }(x)$ is a cylinder function of the order%
\begin{equation}
\nu =\nu (D)=\sqrt{D^{2}/4-D(D+1)\xi +m^{2}a^{2}},  \label{nu}
\end{equation}%
and%
\begin{equation}
\omega =\sqrt{\lambda ^{2}+k^{2}},\;k^{2}=\mathbf{k}_{p}^{2}+\mathbf{k}%
_{q}^{2}.  \label{lamb}
\end{equation}%
Now, the set of quantum numbers is specified as $\sigma =(\mathbf{k}_{p},%
\mathbf{n}_{q},\lambda )$, with $\mathbf{n}_{q}=(n_{p+1},\ldots ,n_{D-1})$.
For imaginary values of $\nu $ AdS spacetime is unstable \cite{Brei82} and
in what follows we assume that the parameter $\nu $ is real.

The factor $C$ in (\ref{fy}) is determined from the normalization condition
\begin{equation}
\int d^{D}x\,\sqrt{|g|}g^{00}\varphi _{\sigma }^{(j)}(x)\varphi _{\sigma
^{\prime }}^{(j^{\prime })\ast }(x)=\frac{1}{2\omega }\delta _{jj^{\prime
}}\delta _{\sigma \sigma ^{\prime }},  \label{Norm}
\end{equation}%
where $\delta _{\sigma \sigma ^{\prime }}$ is understood as Kronecker delta
for discrete components of $\sigma $ and as the Dirac delta function for the
continuous ones. For $\nu \geqslant 1$, from the normalizability of the mode
functions it follows that one should take $Z_{\nu }(\lambda z)=J_{\nu
}(\lambda z)$ being $J_{\nu }(x)$ the Bessel function of the first kind. For
the solution with the Neumann function $Y_{\nu }(x)$, the normalization
integral in (\ref{Norm}) diverges on the AdS boundary $z=0$. Note that in
AdS/CFT correspondence normalizable and non-normalizable modes are dual to
states and sources, respectively. For $0\leqslant \nu <1$, in order to
uniquely define the mode functions for the quantization procedure, it is
necessary to specify boundary conditions at the AdS boundary \cite%
{Brei82,Avis78}. The Dirichlet and Neumann boundary conditions are the most
frequently used ones. The general class of allowed boundary conditions of
the Robin type, has been specified in \cite{Ishi04} on the base of general
analysis in \cite{Ishi03}. In what follows, for the modes with $0\leqslant
\nu <1$, we shall choose $Z_{\nu }(\lambda z)=J_{\nu }(\lambda z)$ that
corresponds to Dirichlet condition on the AdS boundary. In this case, from (%
\ref{Norm}) for the coefficient in (\ref{fy}) one gets%
\begin{equation}
|C|^{2}=\frac{a^{1-D}\lambda }{2(2\pi )^{p}\omega V_{q}},  \label{C2}
\end{equation}%
where $V_{q}=L_{p+1}\cdots L_{D-1}$ is the volume of the compact subspace.

Substituting the mode functions into (\ref{G11}), for the Hadamard function
we obtain%
\begin{equation}
G^{(1)}(x,x^{\prime })=\frac{a^{1-D}(zz^{\prime })^{D/2}}{(2\pi )^{p}V_{q}}%
\sum_{\mathbf{n}_{q}\in \mathbf{Z}^{q}}e^{i\mathbf{k}_{q}\cdot \Delta
\mathbf{x}_{q}}\int d\mathbf{k}_{p}\,e^{i\mathbf{k}_{p}\cdot \Delta \mathbf{x%
}_{p}}\int_{0}^{\infty }d\lambda \,\frac{\lambda }{\omega }J_{\nu }(\lambda
z)J_{\nu }(\lambda z^{\prime })\cos (\omega \Delta t),  \label{G12}
\end{equation}%
where $\Delta \mathbf{x}_{q}=\mathbf{x}_{q}-\mathbf{x}_{q}^{\prime }$, $%
\Delta \mathbf{x}_{p}=\mathbf{x}_{p}-\mathbf{x}_{p}^{\prime }$, $\Delta
t=t-t^{\prime }$. By using the transformations described in appendix, this
function is presented in the final form%
\begin{equation}
G^{(1)}(x,x^{\prime })=\frac{2a^{1-D}}{(2\pi )^{(D+1)/2}}\sum_{\mathbf{n}%
_{q}\in \mathbf{Z}^{q}}e^{i\tilde{\mathbf{\alpha }}\cdot \mathbf{n}%
_{q}}q_{\nu -1/2}^{(D-1)/2}(v_{\mathbf{n}_{q}}),  \label{G15}
\end{equation}%
where we have defined%
\begin{equation}
v_{\mathbf{n}_{q}}=1+\frac{(\Delta \mathbf{x}_{p})^{2}+\sum_{i=p+1}^{D-1}%
\left( \Delta x^{i}-L_{i}n_{i}\right) ^{2}+(\Delta z)^{2}-(\Delta t)^{2}}{%
2zz^{\prime }},  \label{vn}
\end{equation}%
with $\Delta z=z-z^{\prime }$. In (\ref{G15}), we have introduced the
function%
\begin{eqnarray}
q_{\alpha }^{\mu }(x) &=&\frac{e^{-i\pi \mu }Q_{\alpha }^{\mu }(x)}{%
(x^{2}-1)^{\mu /2}}  \notag \\
&=&\frac{\sqrt{\pi }\Gamma (\alpha +\mu +1)}{2^{\alpha +1}\Gamma (\alpha
+3/2)x^{\alpha +\mu +1}}F\left( \frac{\alpha +\mu }{2}+1,\frac{\alpha +\mu +1%
}{2};\alpha +\frac{3}{2};\frac{1}{x^{2}}\right) ,  \label{qmu}
\end{eqnarray}%
being $Q_{\alpha }^{\mu }(x)$ the associated Legendre function of the second
kind and $F(a,b;c;u)$ is the hypergeometric function. By using the
recurrence relations for the associated Legendre functions we can obtain the
following relation%
\begin{equation}
\partial _{x}q_{\alpha }^{\mu }(x)=-q_{\alpha }^{\mu +1}(x).  \label{Relqmu}
\end{equation}%
Note that in (\ref{G15}) the term with $\mathbf{n}_{q}=0$ corresponds to the
Hadamard function in AdS spacetime with trivial topology of the subspace
perpendicular to the $y$-axis. The remaining terms are induced by the
compactification. The quantity $v_{\mathbf{n}_{q}}$ with $\mathbf{n}_{q}=0$
is related to the invariant distance $d(x,x^{\prime })$ by $v_{\mathbf{0}%
}=\cosh (d(x,x^{\prime }))$.

For a conformally coupled massless field one has $\nu =1/2$ and the
hypergeometric function in (\ref{G15}) is expressed as
\begin{equation}
F\left( \frac{\nu +D/2}{2},\frac{\nu +D/2+1}{2};\nu +1;1/v_{\mathbf{n}%
_{q}}^{2}\right) =\frac{v_{\mathbf{n}_{q}}}{1-D}\sum_{j=-1,1}j(1+j/v_{%
\mathbf{n}_{q}})^{(1-D)/2}.  \label{Hyp}
\end{equation}%
For the Hadamard function this gives%
\begin{equation}
G^{(1)}(x,x^{\prime })=(zz^{\prime }/a^{2})^{(D-1)/2}G_{M}^{(1)}(x,x^{\prime
}),  \label{G1cc}
\end{equation}%
where%
\begin{equation}
G_{M}^{(1)}(x,x^{\prime })=\frac{\Gamma ((D-1)/2)}{2\pi ^{(D+1)/2}}\sum_{%
\mathbf{n}_{q}\in \mathbf{Z}^{q}}e^{i\tilde{\mathbf{\alpha }}\cdot \mathbf{n}%
_{q}}\left[ (\Delta x_{\mathbf{n}_{q}}^{(-)})^{1-D}-(\Delta x_{\mathbf{n}%
_{q}}^{(+)})^{1-D}\right] ,  \label{G1M}
\end{equation}%
with the notation%
\begin{equation}
(\Delta x_{\mathbf{n}_{q}}^{(\pm )})^{2}=(\Delta \mathbf{x}%
_{p})^{2}+\sum_{i=p+1}^{D-1}\left( \Delta x^{i}-L_{i}n_{i}\right) ^{2}+(z\pm
z^{\prime })^{2}-(\Delta t)^{2}.  \label{Delxpm}
\end{equation}%
It can be shown that the function (\ref{G1M}) coincides with the Hadamard
function for a massless scalar field in $(D+1)$-dimensional Minkowski
spacitme with uncompactified dimensions $(x^{1},\ldots ,x^{p},z)$ and
compactified dimensions $(x^{p+1},\ldots ,x^{D-1})$, in the presence of a
planar boundary at $z=0$ on which the field obeys Dirichlet boundary
condition. The formula (\ref{G1cc}) is the standard conformal relation
between the conformally connected problems \cite{Birr82}. For a conformally
coupled massless field the problem at hand is conformally related to the
corresponding problem in Minkowski spacetime with a planar boundary. The
presence of the latter is a consequence of the boundary condition we have
imposed on the AdS boundary. Note that the part in the function $%
G_{M}^{(1)}(x,x^{\prime })$ with the first term in the square brackets of (%
\ref{G1M}), is the Hadamard function in boundary-free Minkowski spacetime
with toroidally compactified dimensions. The remaining part is induced by
the boundary.

\section{Vacuum current density}

\label{sec:Current}

With the Hadamard function from (\ref{G15}), the VEV of the current density
is evaluated by making use of formula (\ref{jmu2}). The charge density and
the components along uncompactified dimensions vanish. For the component
along the $l$th compact dimension, by using the relation (\ref{Relqmu}), one
gets%
\begin{equation}
\langle j^{l}\rangle =\frac{4ea^{-1-D}L_{l}}{(2\pi )^{(D+1)/2}}%
\sum_{n_{l}=1}^{\infty }n_{l}\sin (\tilde{\alpha}_{l}n_{l})\sum_{\mathbf{n}%
_{q-1}}\,\cos (\tilde{\mathbf{\alpha }}_{q-1}\cdot \mathbf{n}_{q-1})q_{\nu
-1/2}^{(D+1)/2}(1+g_{\mathbf{n}_{q}}^{2}/(2z^{2})),  \label{jl}
\end{equation}%
where $\mathbf{n}_{q-1}=(n_{p+1},\ldots ,n_{l-1},n_{l+1},\ldots ,n_{D-1})$,
\begin{equation}
\tilde{\mathbf{\alpha }}_{q-1}\cdot \mathbf{n}_{q-1}=\sum_{i=1,\neq l}^{D-1}%
\tilde{\alpha}_{i}n_{i},\;g_{\mathbf{n}_{q}}=\left(
\sum_{i=p+1}^{D-1}n_{i}^{2}L_{i}^{2}\right) ^{1/2},  \label{alfq-1}
\end{equation}%
and the summation goes over all values of $-\infty <n_{i}<+\infty $, $i\neq l
$. Note that in (\ref{jl}) instead of $\cos (\tilde{\mathbf{\alpha }}%
_{q-1}\cdot \mathbf{n}_{q-1})$ we can also write $\prod\nolimits_{i=1,\neq
l}^{D-1}\cos (\tilde{\alpha}_{i}n_{i})$. As is seen, the current density
along the $l$th compact dimension is an odd periodic function of the phase $%
\tilde{\alpha}_{l}$ and an even periodic function of the phases $\tilde{%
\alpha}_{i}$, $i\neq l$. In both cases the period is equal to $2\pi $. In
particular, the current density is a periodic function of the magnetic
fluxes with the period equal to the flux quantum $2\pi /|e|$. In the absence
of the gauge field, the current density along the $l$th compact dimension
vanishes for untwisted and twisted fields along that direction. Of course,
the latter is a direct consequence of the problem symmetry under the
reflection $x^{l}\rightarrow -x^{l}$ in these special cases.

The charge flux through the $(D-1)$-dimensional spatial hypersurface
$x^{l}=\mathrm{const}$ is determined by the quantity $n_{l}\langle
j^{l}\rangle $, where $n_{l}=a/z$ is the corresponding normal. From
(\ref{jl}) we see that this quantity depends on the coordinate
lengths of the compact dimensions $L_{i}$ in the form of the ratio
$L_{i}/z$. For a given $z$, the latter is the proper length of the
compact dimension, $L_{(p)i}=aL_{i}/z$, measured in units of the AdS
curvature radius $a$. This property is a consequence of the maximal
symmetry of AdS spacetime. Because $\langle j^{z}\rangle =0$ and the VEV (%
\ref{jl}) depends on the coordinate $z$ only, the covariant conservation
equation $\nabla _{l}\langle j^{l}\rangle =0$ is satisfied trivially.

The function $q_{\alpha }^{\mu }(x)$ in (\ref{jl}) is defined by (\ref{qmu}%
). By using the relation (\ref{Relqmu}), it can also be expressed in the form%
\begin{equation}
q_{\nu -1/2}^{(D+1)/2}(x)=(-1)^{n}\sqrt{\frac{\pi }{2}}\partial _{x}^{n}%
\frac{(x+\sqrt{x^{2}-1})^{-\nu }}{\sqrt{x^{2}-1}},  \label{qev}
\end{equation}%
for $D=2n$, $n=1,2,\ldots $, and in the form%
\begin{equation}
q_{\nu -1/2}^{(D+1)/2}(x)=(-1)^{n}\partial _{x}^{n}Q_{\nu -1/2}(x),
\label{qod}
\end{equation}%
for $D=2n-1$. In what follows we shall need the asymptotic expressions in
different limiting cases. For $x\gg 1$ one has%
\begin{equation}
q_{\nu -1/2}^{(D+1)/2}(x)\approx \frac{\sqrt{\pi }\Gamma (\nu +D/2+1)}{%
2^{\nu +1/2}\Gamma (\nu +1)x^{D/2+\nu +1}}.  \label{fnul}
\end{equation}%
If $x$ is close to 1, $x-1\ll 1$, by using the asymptotic formula for the
hypergeometric function, we get
\begin{equation}
q_{\nu -1/2}^{(D+1)/2}(x)\approx \frac{\Gamma ((D+1)/2)}{2\left( x-1\right)
^{(D+1)/2}}.  \label{fDxto1}
\end{equation}%
Finally, for large values of the order, $\nu \gg 1$, one has%
\begin{equation}
q_{\nu -1/2}^{(D+1)/2}(x)\approx \sqrt{\frac{\pi }{2}}\frac{\nu ^{D/2}(x+%
\sqrt{x^{2}-1})^{-\nu }}{(x^{2}-1)^{(D+2)/4}}.  \label{qulargenu}
\end{equation}

An alternative expression for the VEV of the current density is obtained by
using the integral representation (\ref{G14}) given in the appendix for the
Hadamard function:%
\begin{equation}
\langle j^{l}\rangle =\frac{2eL_{l}}{(2\pi )^{D/2}a^{D+1}}%
\sum_{n_{l}=1}^{\infty }n_{l}\sin (\tilde{\alpha}_{l}n_{l})\sum_{\mathbf{n}%
_{q-1}}\,\cos (\tilde{\mathbf{\alpha }}_{q-1}\cdot \mathbf{n}%
_{q-1})\int_{0}^{\infty }dx\,\frac{x^{D/2}I_{\nu }(x)}{e^{x[1+g_{\mathbf{n}%
_{q}}^{2}/(2z^{2})]}}.  \label{jlalt}
\end{equation}%
This representation is useful in the discussion of limiting cases.

For a conformally coupled massless field one has $\nu =1/2$ and for the
current density we get
\begin{equation}
\langle j^{l}\rangle =(z/a)^{D+1}\langle j^{l}\rangle _{M}^{(b)},
\label{jlcc}
\end{equation}%
where%
\begin{eqnarray}
\langle j^{l}\rangle _{M}^{(b)} &=&2eL_{l}\frac{\Gamma ((D+1)/2)}{\pi
^{(D+1)/2}}\sum_{n_{l}=1}^{\infty }n_{l}\sin (\tilde{\alpha}_{l}n_{l})\sum_{%
\mathbf{n}_{q-1}}\,\cos (\tilde{\mathbf{\alpha }}_{q-1}\cdot \mathbf{n}%
_{q-1})  \notag \\
&&\times \left[ \frac{1}{g_{\mathbf{n}_{q}}^{D+1}}-\frac{1}{(g_{\mathbf{n}%
_{q}}^{2}+4z^{2})^{(D+1)/2}}\right] ,  \label{jlM}
\end{eqnarray}%
is the current density for a massless scalar field in Minkowski
spacetime with toroidally compactified dimensions, in the presence
of Dirichlet boundary at $z=0$. The part with the first term in
figure braces corresponds to the VEV in the boundary-free Minkowski
spacetime. It is obtained from the general result of \cite{Beze13b}
in the zero mass limit.

Now we turn to the investigation of the current density in various
asymptotic regions of the parameters. First let us consider the Minkowskian
limit corresponding to $a\rightarrow \infty $ for a fixed value of the
coordinate $y$. In this limit $\nu \approx ma\gg 1$ and for the coordinate $z
$ one has $z\approx a+y$. Introducing in (\ref{jlalt}) a new integration
variable $x=\nu u$ we use the uniform asymptotic expansion for the function $%
I_{\nu }(\nu u)$ for large values of the order \cite{Watson}. The dominant
contribution to the integral comes from large values of $u$ and, after the
integration, to the leading order we get
\begin{equation}
\langle j^{l}\rangle \approx \frac{4eL_{l}m^{(D+1)/2}}{(2\pi )^{(D+1)/2}}%
\sum_{n_{l}=1}^{\infty }n_{l}\sin (\tilde{\alpha}_{l}n_{l})\sum_{\mathbf{n}%
_{q-1}}\,\cos (\tilde{\mathbf{\alpha }}_{q-1}\cdot \mathbf{n}_{q-1})\frac{%
K_{(D+1)/2}(mg_{\mathbf{n}_{q}})}{g_{\mathbf{n}_{q}}^{(D+1)/2}},
\label{jlMinkLim}
\end{equation}%
being $K_{(D+1)/2}(x)$ the MacDonald function. The expression in the
right-hand side coincides with the corresponding result for a scalar field
in Minkowski spacetime with toroidally compactified dimensions obtained in
\cite{Beze13b}.

If the proper length of the one of the compact dimensions, say $x^{i}$, $%
i\neq l$, is large compared with the AdS curvature radius, $L_{i}/z\gg 1$,
the dominant contribution into (\ref{jl}) comes from the $n_{i}=0$ term and
the contribution of the remaining terms is suppressed by the factor $%
(z/L_{i})^{D+2\nu +2}$. To the leading order, we obtain the current density
for the topology $R^{p+1}\times (S^{1})^{q-1}$ with the uncompactified
direction $x^{i}$. In the opposite limit corresponding to small proper
length of the dimension $x^{i}$, $L_{i}/z\ll 1$, the dominant contribution
comes from large values of $n_{i}$ and for the corresponding series in (\ref%
{jlalt}) we use the relation%
\begin{equation}
\sum_{n_{i}=-\infty }^{+\infty }\cos (\tilde{\alpha}%
_{i}n_{i})e^{-xL_{i}^{2}n_{i}^{2}/(2z^{2})}\approx \frac{2z}{L_{i}}\sqrt{%
\frac{\pi }{2x}}e^{-(\sigma _{i}z/L_{i})^{2}/(2x)},  \label{Rel4}
\end{equation}%
with $\sigma _{i}=\mathrm{min}(\tilde{\alpha}_{i},2\pi -\tilde{\alpha}_{i})$%
, $0\leqslant \tilde{\alpha}_{i}<2\pi $. The behavior of the current density
depends crucially on whether the phase $\tilde{\alpha}_{i}$ is zero or not.
For $\tilde{\alpha}_{i}=0$, to the leading order, for the combination $%
(aL_{i}/z)\langle j^{l}\rangle $ we obtain the expression which coincides
with the formula for $\langle j^{l}\rangle $ in $D$-dimensional AdS
spacetime, obtained from the geometry under consideration by excluding the
dimension $x^{i}$, with the replacement $\nu (D-1)\rightarrow \nu (D)$,
where the function $\nu (D)$ is defined by (\ref{nu}). For $\tilde{\alpha}%
_{i}\neq 0$, after the substitution (\ref{Rel4}), the dominant contribution
in the integral of (\ref{jlalt}) comes from the values $x\gtrsim z/L_{i}$.
Replacing the modified Bessel function by its asymptotic expression for
large values of the argument, we can see that the dominant contribution
comes from the term $n_{l}=1$, $n_{r}=0$, $r\neq i,l$, and to the leading
order one gets%
\begin{equation}
\langle j^{l}\rangle \approx \frac{2eL_{l}\sigma _{i}^{(D-1)/2}\sin (\tilde{%
\alpha}_{l})}{(2\pi )^{(D-1)/2}a^{D+1}}\frac{z^{D+1}e^{-L_{l}\sigma
_{i}/L_{i}}}{(L_{i}L_{l})^{(D+1)/2}}.  \label{jlLismall}
\end{equation}%
In this case the current density is exponentially small.

Now let us consider the behavior of the current $\langle j^{l}\rangle $ for
large and small values of $L_{l}$. For large values of the proper length
compared with the AdS curvature radius, $L_{l}/z\gg 1$, the argument of the
function $q_{\nu -1/2}^{(D+1)/2}(x)$ in (\ref{jl}) is large and by using (%
\ref{fnul}) in the leading order we find%
\begin{equation}
\langle j^{l}\rangle \approx \frac{eL_{l}(2z^{2})^{D/2+\nu +1}}{2^{D/2+\nu
-1}a^{D+1}\pi ^{D/2}}\frac{\Gamma (\nu +D/2+1)}{\Gamma (\nu +1)}%
\sum_{n_{l}=1}^{\infty }n_{l}\sin (\tilde{\alpha}_{l}n_{l})\sum_{\mathbf{n}%
_{q-1}}\,\frac{\cos (\tilde{\mathbf{\alpha }}_{q-1}\cdot \mathbf{n}_{q-1})}{%
g_{\mathbf{n}_{q}}^{D+2\nu +2}}.  \label{jlLllarge}
\end{equation}%
The dominant contribution comes from large values of $n_{i}$, $i\neq l$, and
we replace the summation $\sum_{\mathbf{n}_{q-1}}$ by the integration. For $%
\tilde{\alpha}_{i}=0$, $i=p+1,\ldots ,D-1$, $i\neq l$, to the leading order
one finds%
\begin{equation}
\langle j^{l}\rangle \approx \frac{4e\Gamma (p/2+\nu +2)}{\pi ^{p/2+1}\Gamma
(\nu +1)a^{D+1}V_{q}}\frac{z^{D+2\nu +2}}{L_{l}^{p+2\nu +2}}%
\sum_{n_{l}=1}^{\infty }\frac{\sin (\tilde{\alpha}_{l}n_{l})}{n_{l}^{p+2\nu
+3}},  \label{jlLllalf0}
\end{equation}%
with the power-law decay as a function of $L_{l}$ for both massless and
massive fields. In this sense, the situation for the AdS bulk is essentially
different from that in the corresponding problem for Minkowski background.
For the latter, in the massless case and for large values of $L_{l}$ the
current density decays as $1/L_{l}^{p}$, whereas for a massive field the
current is suppressed exponentially, by the factor $e^{-mL_{l}}$. This shows
that the influence of the background gravitational field on the VEV is
crucial. If $L_{l}/z\gg 1$ and at least one of the phases $\tilde{\alpha}_{i}
$, $i\neq l$, is not equal to zero, the dominant contribution comes from the
$n_{l}=1$ term and one gets%
\begin{equation}
\langle j^{l}\rangle \approx \frac{2ea^{-1-D}}{\pi ^{(p+1)/2}}\frac{\sin (%
\tilde{\alpha}_{l})z^{D+2\nu +2}}{\Gamma (\nu +1)V_{q}e^{\beta _{q-1}L_{l}}}%
\frac{\beta _{q-1}^{(p+3)/2+\nu }}{(2L_{l})^{(p+1)/2+\nu }},
\label{jlLllalf}
\end{equation}%
where $\beta _{q-1}=(\sum_{i=p+1,\neq l}^{D-1}\tilde{\alpha}%
_{i}^{2}/L_{i}^{2})^{1/2}$. In this case the current density, as a function
of $L_{l}$, decays exponentially.

For small values of $L_{l}$, $L_{l}/z\ll 1$, the dominant contribution to
the current density comes from the term with $\mathbf{n}_{q-1}=0$ and to the
leading order we obtain%
\begin{equation}
\langle j^{l}\rangle \approx \frac{2e\Gamma ((D+1)/2)}{\pi
^{(D+1)/2}(a/z)^{D+1}L_{l}^{D}}\sum_{n_{l}=1}^{\infty }\frac{\sin (\tilde{%
\alpha}_{l}n_{l})}{n_{l}^{D}}.  \label{jlLlsm}
\end{equation}%
In this limit, the contribution of the terms with $\mathbf{n}_{q-1}\neq 0$
is suppressed by the factor $e^{-\sigma _{l}|\mathbf{L}\cdot \mathbf{n}%
_{q-1}|/L_{l}}$, where $\sigma _{l}=\mathrm{min}(\tilde{\alpha}_{l},2\pi -%
\tilde{\alpha}_{l})$, $0<\tilde{\alpha}_{l}<2\pi $. The expression in the
right-hand side of (\ref{jlLlsm}), multiplied by $(a/z)^{D+1}$, coincides
with the VEV of the current density for a massless scalar field in $(D+1)$%
-dimensional Minkowski spacetime compactified along the direction $x^{l}$ to
the circle with the length $L_{l}$.

Now let us investigate the behavior of the current density near the AdS
boundary and near the horizon for fixed values of the lengths $L_{i}$. Near
the boundary, $z\rightarrow 0$, the argument of the function $q_{\nu
-1/2}^{(D+1)/2}(x)$ in (\ref{jl}) is large and to the leading order we find
\begin{equation}
\langle j^{l}\rangle \approx \frac{4eL_{l}\Gamma (\nu +D/2+1)}{\pi
^{D/2}\Gamma (\nu +1)a^{D+1}}z^{D+2\nu +2}\sum_{n_{l}=1}^{\infty }n_{l}\sin (%
\tilde{\alpha}_{l}n_{l})\sum_{\mathbf{n}_{q-1}}\,\frac{\cos (\tilde{\mathbf{%
\alpha }}_{q-1}\cdot \mathbf{n}_{q-1})}{g_{\mathbf{n}_{q}}^{D+2\nu +2}}.
\label{jlbound}
\end{equation}%
Hence, the current density vanishes on the boundary. Near the horizon one
has $z\rightarrow \infty $, the argument of the function $q_{\nu
-1/2}^{(D+1)/2}(x)$ is close to 1 and we use the asymptotic expression (\ref%
{fDxto1}). This gives,%
\begin{equation}
\langle j^{l}\rangle \approx (z/a)^{D+1}\langle j^{l}\rangle _{M},
\label{jlhor}
\end{equation}%
where
\begin{equation}
\langle j^{l}\rangle _{M}=2eL_{l}\frac{\Gamma ((D+1)/2)}{\pi ^{(D+1)/2}}%
\sum_{n_{l}=1}^{\infty }n_{l}\sin (\tilde{\alpha}_{l}n_{l})\sum_{\mathbf{n}%
_{q-1}}\,\frac{\cos (\tilde{\mathbf{\alpha }}_{q-1}\cdot \mathbf{n}_{q-1})}{%
g_{\mathbf{n}_{q}}^{D+1}},  \label{jlMm0}
\end{equation}%
is the corresponding current density in Minkowski spacetime for a massless
scalar field.

And finally, for large values of the mass, $ma\gg 1$, by using the
asymptotic expression (\ref{qulargenu}) we see that the dominant
contribution in (\ref{jl}) comes from the term with $n_{i}=0$, $i\neq l$,
and $n_{l}=1$. To the leading order we get%
\begin{equation}
\langle j^{l}\rangle \approx \frac{2eL_{l}\sin (\tilde{\alpha}_{l})}{(2\pi
)^{D/2}a^{D+1}}\frac{(ma)^{D/2}u^{-D/2-1}(1+u^{2}/4)^{-(D+2)/4}}{(1+u^{2}/2+u%
\sqrt{1+u^{2}/4})^{ma}},  \label{jllargem}
\end{equation}%
with $u=L_{l}/z$. As we could expect, for large values of the mass we have
an exponential suppression. The suppression is stronger near the AdS
boundary.

\begin{figure}[tbph]
\begin{center}
\begin{tabular}{cc}
\epsfig{figure=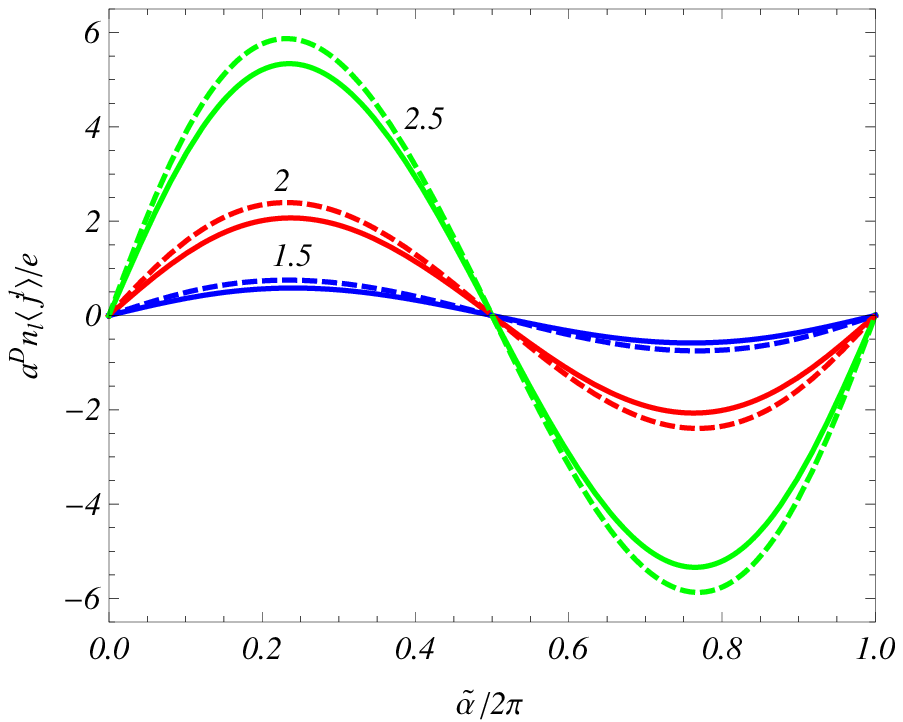,width=7.cm,height=5.5cm} & \quad %
\epsfig{figure=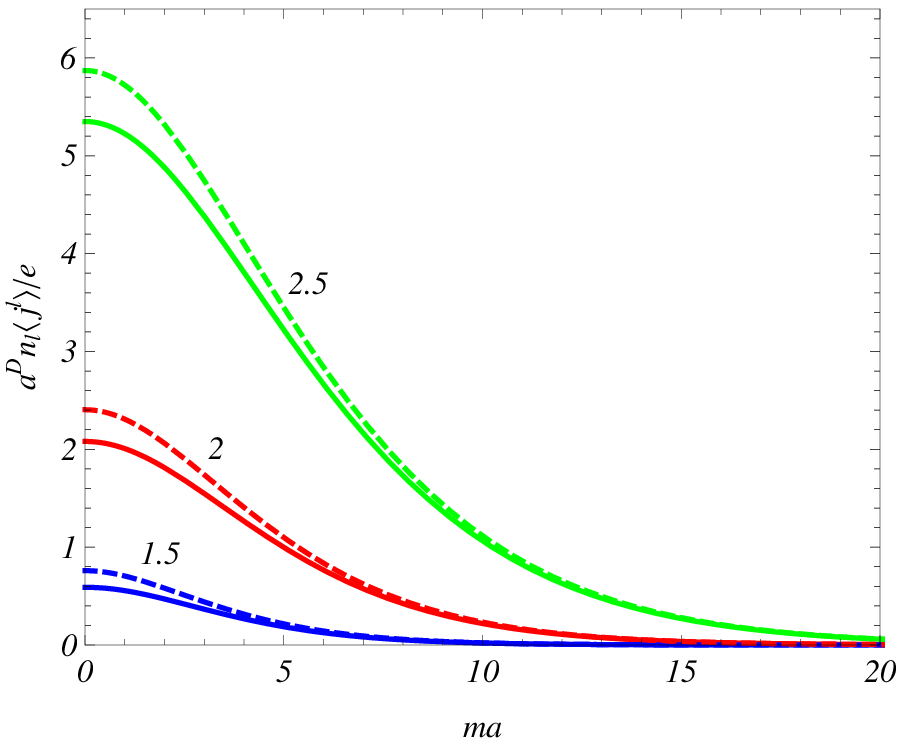,width=7.cm,height=5.5cm}%
\end{tabular}%
\end{center}
\caption{The quantity $a^{D}n_{l}\langle j^{l}\rangle /e$ as a function of $%
\tilde{\protect\alpha}$ (left panel) and as a function of $ma$ (right panel)
for $D=4$ AdS space with a single compact dimension $x^{D-1}$. The numbers
near the curves correspond to the values of the ratio $z/L$ and the
full/dashed curves are for minimally/conformally coupled fields. For the
graphs on the left panel $ma=0.5$ and for the right panel $\tilde{\protect%
\alpha}=\protect\pi /2$.}
\label{fig2}
\end{figure}

In the special case of a single compact dimension (dimension $x^{D-1}$) with
the length $L_{D-1}=L$ from the general formula one finds
\begin{equation}
\langle j^{D-1}\rangle =\frac{4ea^{-1-D}L}{(2\pi )^{(D+1)/2}}%
\sum_{n=1}^{\infty }n\sin (\tilde{\alpha}n)q_{\nu
-1/2}^{(D+1)/2}(1+n^{2}L^{2}/(2z^{2})),  \label{jD-1}
\end{equation}%
where $\tilde{\alpha}=\tilde{\alpha}_{D-1}$. For this special case with $D=4$%
, in figure \ref{fig2} we have displayed the quantity
$a^{D}n_{l}\langle j^{l}\rangle /e$ as a function of the phase in
the periodicity condition (left panel) and as a function of the mass
(right panel). The numbers near the curves correspond to the values
of the ratio $z/L$ and the full/dashed curves are for
minimally/conformally coupled fields. On the left panel, the
graphs are plotted for $ma=0.5$ and for the right panel we have taken $%
\tilde{\alpha}=\pi /2$.

\begin{figure}[tbph]
\begin{center}
\epsfig{figure=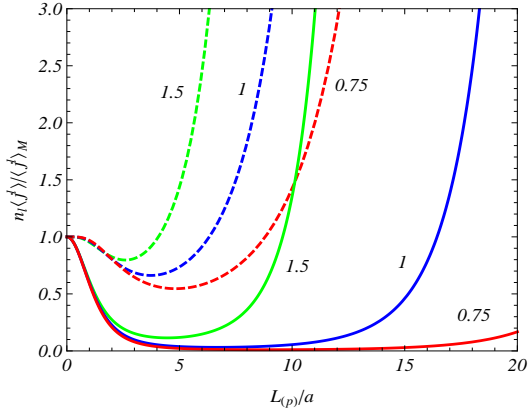,width=7.cm,height=5.5cm}
\end{center}
\caption{The ratio of the current densities in AdS and Minkowski backgrounds
as a function of the proper length of the compact dimension. The numbers
near the curves correspond to the values of $ma$. The full and dashed curves
are for minimally and conformally coupled fields. For the phase in the
periodicity condition we have taken $\tilde{\protect\alpha}=\protect\pi /2$.}
\label{fig3}
\end{figure}

For the same model with $D=4$, in figure \ref{fig3} we have plotted the
ratio of the current densities in AdS and Minkowski bulks for the same
proper lengths of the compact dimension, $L_{(p)}$, as a function of the
proper length measured in units of the AdS curvature radius. The current
density in Minkowski bulk is given by the right-hand side of (\ref{jlMinkLim}%
), specified to the special case under consideration. The graphs are plotted
for $\tilde{\alpha}=\pi /2$ and the numbers near the curves are the
corresponding values of the parameter $ma$ (mass measured in units of the
AdS energy scale). As before the full and dashed curves correspond to
minimally and conformally coupled fields. Note that in the case of the AdS
bulk one has $L_{(p)}=aL/z$, with $L$ being the coordinate length. In the
Minkowski spacetime with compact dimension the proper and coordinate lengths
coincide. From figure \ref{fig3} we see the feature already described before
on the base of the asymptotic analysis: for a massive field and for large
values of the proper length the decay of the current density in the
Minkowski bulk is stronger than that for AdS background.

\section{Conclusion}

\label{sec:Conc}

In the present paper we have investigated combined effects of
geometry and topology on the VEV of the current density for a
charged scalar field. In order to have an exactly solvable problem
we have taken highly symmetric background geometry corresponding to
a slice of AdS spacetime with a toroidally compactified subspace.
Currently, the AdS spacetime is among the most popular backgrounds
in gravitational physics and appears in a number of contexts. The
information on the vacuum fluctuations is encoded in two-point
functions and, as the first step, we have evaluated the Hadamard
function for general values of the phases in the periodicity
conditions obeyed by the field operator along compact dimensions.
Additionally, the presence of a constant gauge field is assumed. The
latter may be excluded by a gauge transformation that leads to the
shift in the phases of periodicity conditions. The shift is
expressed in terms of the ratio of the magnetic flux enclosed by
compact dimension to the flux quantum. The Hadamard function is
expressed in the form (\ref{G15}), where the part corresponding to
AdS geometry without compactification is explicitly separated and is
presented by the term with $\mathbf{n}_{q}=0$. Because the toroidal
compactification does not change the local geometry, in this way the
renormalization of the VEVs for physical quantities bilinear in the
field is reduced to the one for the uncompactified AdS spacetime.

With a given Hadamard function, the VEV of the current density is evaluated
by using the formula (\ref{jmu2}). The vacuum charge density and the
components of the current along uncompactified dimensions vanish and the
current density along $l$th compact dimension is given by the expression (%
\ref{jl}). The latter is an odd periodic function of the phase $\tilde{\alpha%
}_{l}$ and an even periodic function of the phases $\tilde{\alpha}_{i}$, $%
i\neq l$. In particular, the current density is a periodic function of the
magnetic fluxes with the period equal to the flux quantum. For a conformally
coupled massless field the current density is conformally related to the
corresponding quantity in $(D+1)$-dimensional Minkowski spacitme with toral
dimensions, in the presence of a planar boundary on which the field obeys
Dirichlet boundary condition. The appearance of the latter is a consequence
of the boundary condition we have imposed on the AdS boundary. As an
additional check, we have shown that in the Minkowskian limit, corresponding
to $a\rightarrow \infty $ for a fixed value of the coordinate $y$, the
result obtained in \cite{Beze13} is recovered.

In order to clarify the dependence of the current density on the values of
the parameters characterizing the geometry and the topology, we have
considered various limiting cases. Near the AdS boundary, the VEV of the
current density behaves as $z^{D+2\nu +2}$ and, hence, it vanishes on the
AdS boundary. For fixed values of the coordinate lengths of compact
dimensions, this region corresponds to large values of the proper lengths.
Near the horizon, corresponding to $z=\infty $, the VEV of the current
density is related to the current density in Minkowski spacetime with toral
dimensions by the relation (\ref{jlhor}) and behaves as $z^{D+1}$. It is
also of interest to consider the behavior for large and small proper lengths
of compact dimensions for a fixed value of $z$. If the proper length of the $%
i$th compact dimension is small, then the behavior of the current density
along the $l$th dimension depends crucially on whether the phase $\tilde{%
\alpha}_{i}$ is zero or not. For the zero value of this phase, in the
leading order, the expression for the quantity $L_{(p)i}\langle j^{l}\rangle
$ coincides with the formula for $\langle j^{l}\rangle $ in $D$-dimensional
AdS spacetime, obtained from the geometry at hand by excluding the dimension
$x^{i}$ and replacing $\nu (D-1)\rightarrow \nu (D)$, where the function $%
\nu (D)$ is defined by (\ref{nu}). For $\tilde{\alpha}_{i}\neq 0$ the
current density $\langle j^{l}\rangle $ is suppressed by the factor $%
e^{-L_{l}\sigma _{i}/L_{i}}$, where $\sigma _{i}$ is defined in the
paragraph after formula (\ref{Rel4}). For large values of the proper length
of the $l$th dimension, compared with the AdS curvature radius, and for zero
values of the phases along the remaining directions, the decay of the VEV $%
\langle j^{l}\rangle $, as a function of $L_{l}$ is power-law, as $%
1/L_{l}^{p+2\nu +2}$ for both massless and massive fields. For massive
fields this behavior is crucially different from that for the Minkowski bulk
where the decay is exponential. If at least one of the phases $\tilde{\alpha}%
_{i}$, $i\neq l$, is not equal to zero, the suppression of the current
density is exponential, given by (\ref{jlLllalf}). For small values of the
proper length of the $l$th dimension, in the leading order, the VEV $\langle
j^{l}\rangle $ does not depend on the phases along other directions and the
current density behaves as $1/L_{l}^{D+1}$.

\section*{Acknowledgments}

E.R.B.M. and A.A.S. thank Conselho Nacional de Desenvolvimento Cient\'{\i}%
fico e Tecnol\'{o}gico (CNPq) for the financial support. A.A.S. was
supported by the State Committee of Science of the Ministry of Education and
Science RA, within the frame of Grant No. SCS 13-1C040.

\appendix

\section{Transformation of the Hadamard function}

\begin{figure}[tbph]
\begin{center}
\epsfig{figure=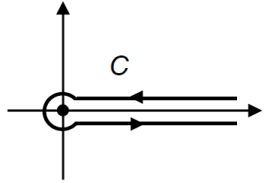,width=3.5cm,height=2.5cm}
\end{center}
\caption{The contour of the integration in the representation (\protect\ref%
{IntRepCos}).}
\label{fig4}
\end{figure}

For the transformation of the expression (\ref{G12}) we use the integral
representation of the Bessel function (see \cite{Watson})
\begin{equation}
J_{\mu }(x)=\frac{(x/2)^{\mu }}{2\pi i}\int_{-\infty }^{(0+)}du\,\frac{%
e^{u-x^{2}/(4u)}}{u^{\mu +1}},  \label{IntRepJ}
\end{equation}%
in which the phase of $u$ increases from $-\pi $ to $\pi $ as $u$ describes
the contour. Taking in this representation $\mu =-1/2$ and changing the
integration variable to $s=ue^{i\pi }$, we can obtain the following
representation%
\begin{equation}
\frac{\cos (\omega \Delta t)}{\omega }=-\frac{1}{2\sqrt{\pi }}\int_{C}\frac{%
ds}{s^{1/2}}\,e^{-\omega ^{2}s+(\Delta t)^{2}/(4s)},  \label{IntRepCos}
\end{equation}%
with the contour of the integration depicted in figure \ref{fig4}. Note that
the integral can also be presented in the form $\int_{C}ds=\int_{c_{\rho
}}ds-2\int_{\rho }^{\infty }ds$, where $c_{\rho }$ is the clockwise oriented
circle of the radius $\rho $, with the center at the origin of the complex $%
s $-plane.

Substituting into (\ref{G12}) and the changing the order of the
integrations, the integral over $\lambda $ is evaluated by using the formula
\cite{Prud86}%
\begin{equation}
\int_{0}^{\infty }d\lambda \,\lambda J_{\nu }(\lambda z)J_{\nu }(\lambda
z^{\prime })e^{-\lambda ^{2}s}=\frac{1}{2s}\frac{I_{\nu }(zz^{\prime }/2s)}{%
e^{(z^{2}+z^{\prime 2})/(4s)}}.  \label{BesInt}
\end{equation}%
For the integral over the momentum along uncompactified dimensions we get%
\begin{equation}
\int d\mathbf{k}_{p}\,e^{i\mathbf{k}_{p}\cdot \Delta \mathbf{x}_{p}-s\mathbf{%
k}_{p}^{2}}=(\pi /s)^{p/2}e^{-|\Delta \mathbf{x}_{p}|^{2}/(4s)}.
\label{kpInt}
\end{equation}%
With these formulas, we can see that, under the condition $(\Delta
t)^{2}<|\Delta \mathbf{x}_{p}|^{2}+z^{2}+z^{\prime 2}$, the part of the $s$%
-integral over the circle $c_{\rho }$ vanishes in the limit $\rho
\rightarrow 0$ and for the remaining integral one gets $\int_{C}ds=-2%
\int_{0}^{\infty }ds$. As a result, the following representation for the
Hadamard function is obtained:%
\begin{equation}
G^{(1)}(x,x^{\prime })=\frac{a^{1-D}(zz^{\prime })^{D/2}}{(4\pi
)^{(p+1)/2}V_{q}}\sum_{\mathbf{n}_{q}\in \mathbf{Z}^{q}}e^{i\mathbf{k}%
_{q}\cdot \Delta \mathbf{x}_{q}}\int_{0}^{\infty }ds\,\frac{I_{\nu
}(zz^{\prime }/2s)}{s^{(p+3)/2}}e^{-s\mathbf{k}_{q}^{2}-[|\Delta \mathbf{x}%
_{p}|^{2}+z^{2}+z^{\prime 2}-(\Delta t)^{2}]/(4s)}.  \label{G13}
\end{equation}

For the further transformation we note that in (\ref{G13}) the series
corresponding to the $l$th compact dimension has the form $%
\sum_{n_{l}=-\infty }^{+\infty }e^{ik_{l}\Delta x^{l}-s^{2}k_{l}^{2}}$. From
the Poisson resummation formula it directly follows that
\begin{equation}
\sum_{n=-\infty }^{+\infty }f(k_{l})=\frac{L_{l}}{2\pi }\sum_{n_{l}=-\infty
}^{+\infty }e^{in_{l}\tilde{\alpha }_{l}}\tilde{f}(n_{l}L_{l}),  \label{Pois}
\end{equation}%
where $\tilde{f}(y)=\int_{-\infty }^{+\infty }dx\,e^{-iyx}f(x)$. From this
formula one gets%
\begin{equation}
\sum_{n_{l}=-\infty }^{+\infty }e^{ik_{l}\Delta x^{l}-sk_{l}^{2}}=\frac{L_{l}%
}{2\sqrt{\pi s}}\sum_{n_{l}=-\infty }^{+\infty }e^{in_{l}\tilde{\alpha }%
_{l}}e^{-(\Delta x^{l}-L_{l}n_{l})^{2}/(4s)}.  \label{SerForm}
\end{equation}%
With this transformation, the Hadamard function is expressed as%
\begin{equation}
G^{(1)}(x,x^{\prime })=\frac{a^{1-D}}{(2\pi )^{D/2}}\sum_{\mathbf{n}_{q}\in
\mathbf{Z}^{q}}e^{i\tilde{\mathbf{\alpha }}\cdot \mathbf{n}%
_{q}}\int_{0}^{\infty }dx\,x^{D/2-1}I_{\nu }(x)e^{-v_{\mathbf{n}_{q}}x},
\label{G14}
\end{equation}%
where%
\begin{equation}
\tilde{\mathbf{\alpha }}=(\tilde{\alpha }_{p+1},\ldots ,\tilde{\alpha }%
_{D-1}),  \label{alfvec}
\end{equation}%
and $v_{\mathbf{n}_{q}}$ is defined by (\ref{vn}). The integral in (\ref{G14}%
) is expressed in terms of the hypergeometric function and we obtain the
representation (\ref{G15}).

\end{document}